\documentclass[preprint,aps,showpacs]{revtex4}
\usepackage{graphicx}

\usepackage{amsmath}
\usepackage{amsxtra}
\usepackage{amstext}
\usepackage{amssymb}
\usepackage{latexsym}
\usepackage{dsfont}

\newcommand{\be}{\begin{equation}}
\newcommand{\ee}{\end{equation}}
\newcommand{\bn}{\begin{eqnarray}}
\newcommand{\en}{\end{eqnarray}}

\begin{document}

\title{What at all is the Higgs of the Standard model and what is the 
origin of families?}

\author{Norma Susana Manko\v c  Bor\v stnik}
\affiliation{Department of Physics, FMF, University of Ljubljana,
Jadranska 19, 1000 Ljubljana}


\begin{abstract}
The {\em standard model} of the elementary particles is built on several assumptions.  
The Higgs is assumed to be a scalar, a boson, with the charges of a fermion (in  
the fundamental representations of the charge groups). 
No explanation is offered  for the existence of families of fermions -- quarks and leptons-- for the 
charges of these family members,  for the appearance of the Yukawas which take care of fermion properties. 
The theory explaining the origin of families predicts that several scalar fields with the 
boson kind of the charges (in the adjoint representations of the charge groups) manifest 
effectively at low energies as the Higgs and the Yukawas. 
\end{abstract}


\maketitle

\section{Introduction}
\label{Introduction}

When the {\em standard model} of the elementary particles and fields was 
proposed more than  35 years ago it offered an elegant new step in understanding 
the origin of fermion and boson fields and the interactions among them. 

It is built on several assumptions, chosen to be in agreement with the data: 
{\bf i.)}   There exist the massless family members - coloured quarks and colourless 
leptons, both "left" and "right handed" (handedness concerns the properties under the 
Lorentz transformations), the left handed members distinguishing from the right 
handed ones in  the weak and hyper charges. 
{\bf ii.)}  There exist the gauge fields to the observed charges of the family members. 
{\bf iii.)} There exists a boson, the Higgs, with a "non zero vacuum expectation value", a scalar  
with the charges of a fermion. Its properties are chosen to "dress" successfully the 
"right handed" family members with the weak and the appropriate "hyper" charge so that they 
manifest the properties of the left handed partners. The Higgs takes  care at the same time 
of  masses of the weak gauge fields $Z_{m}$ and $W^{\pm}_{m}$. 
{\bf iv.)}  There exist the families of fermions.
{\bf v.)}   There exist the Yukawa couplings,  distinguishing among  family members ($u$ and $d$ quarks, 
$e$  and $\nu$ leptons) to ensure right properties of families 
of fermions, that is of their masses and decay properties (mixing matrices).

The properties of fermions and bosons as assumed by the {\it standard model} are presented in 
tables~\ref{Table I.},\ref{Table II.},\ref{Table III.}.

\begin{table}
\begin{center}
\begin{tabular}{|r c c c c r|}
\hline
$\alpha$& handedness &  weak charge & hyper charge  & colour charge& elm  charge \\
%
name         &$ -4i S^{03} S^{12}$ &$ \tau^{13}$  &$ Y$          &               &$Q$\\
\hline
$u^{i}_{L}$  &  left  handed ($-1$)&$ \frac{1}{2}$&$ \frac{1}{6}$& colour triplet&$\frac{2}{3}$\\
$d^{i}_{L}$  &  left  handed ($-1$)&$-\frac{1}{2}$&$ \frac{1}{6}$& colour triplet&$-\frac{1}{3}$\\
$\nu^{i}_{L}$&  left  handed ($-1$)&$ \frac{1}{2}$&$-\frac{1}{2}$&  colourless   &$0$  \\
$e^{i}_{L}  $&  left  handed ($-1$)&$-\frac{1}{2}$&$-\frac{1}{2}$&  colourless   &$-1$ \\
\hline
$u^{i}_{R}  $& right  handed ($ 1$)& weakless     &$ \frac{2}{3}$& colour triplet&$\frac{ 2}{3}$\\
$d^{i}_{R}  $& right  handed ($ 1$)& weakless     &$-\frac{1}{3}$& colour triplet&$-\frac{1}{3}$\\
$\nu^{i}_{R}$& right  handed ($ 1$)& weakless     &$ 0          $&  colourless   &$0$           \\
$e^{i}_{R}  $& right  handed ($ 1$)& weakless     &$ -1         $&  colourless   &$-1$ \\
\hline
\end{tabular}
  \end{center}
\caption{\label{Table I.} The {\it standard model} assumes that there are before the electroweak phase 
transition three ($i=1,2,3$) so far observed massless families of quarks and leptons.  Each family 
contains the left handed weak charged  and the right handed weak chargeless quarks,   belonging  
to the colour triplet $(1/2,1/(2\sqrt{3}))$, $(-1/2,1/(2\sqrt{3}))$, $(0,-1/(\sqrt{3})) $ and the 
colourless left handed weak charged  and the right handed weak chargeless leptons, if in this  
tiny extension of the {\it standard model}   the right handed $\nu$ is added. 
Originally $\nu^{i}_{R}$ were excluded since no massless $\nu$ were observed  and in  the 
{\it standard model} assumption  all the quantum numbers 
of $\nu_R$   are zero. 
$\tau^{13}$ defines the third component of the weak charge, $Y$ the hyper charge,  $Q= Y + \tau^{13}$ is the 
electromagnetic charge. 
} 
\end{table}

\begin{table}
\begin{center}
\begin{tabular}{|r c c c c r|}
\hline
name & handedness    &  weak  charge    & hyper charge &  colour  charge & elm  charge\\
\hline
  hyper photon       &$ 0$              &$ 0$     &  $0$      & colourless  &$0$\\
  weak bosons        &$ 0$              & triplet &  $0$      & colourless  & triplet \\
  gluons             &$ 0$              &$0$      &  $0$      & colour octet&$0$ \\
\hline
\end{tabular}
  \end{center}
\caption{\label{Table II.} The {\it standard model} assumes that there are before the electroweak phase 
transition three massless vector fields, the gauge fields of the three charges - 
the hyper charge ($Y$), the weak charge ($\vec{\tau}^1$) and the colour charge $(\vec{\tau}^3)$, respectively. 
They all are vectors in $d=(1+3)$, carrying the corresponding charges in the adjoint 
representations.  $Q= \tau^{13} + Y$. 
} 
\end{table}

\begin{table}
\begin{center}
\begin{tabular}{|r c c c c r|}
\hline
name       & handedness    &   weak charge  &  hyper charge  &  colour  charge  &  elm charge\\
\hline
Higgs$_{u}$&$ 0$           &$  \frac{1}{2}$&$ \frac{1}{2}  $& colourless&$  1$\\
Higgs$_{d}$&$ 0$           &$- \frac{1}{2}$&$ \frac{1}{2}  $& colourless&$  0$\\                                        
\hline
\end{tabular}
  \end{center}
\caption{\label{Table III.} The {\it standard model} assumes that there is before the electroweak phase 
transition the scalar field Higgs,  a boson, which  carries the hyper charge ($Y$) and the weak charge 
($\vec{\tau}^1$) in the fundamental (spinor) representations of the 
charge groups. It contributes to the phase transition by gaining a non zero "vacuum 
expectation value" of that component which has the electromagnetic charge ($Q= \tau^{13} + Y$) 
equal to zero. Correspondingly it changes properties of the vacuum.
The Higgs "dresses" right handed $(d^{i}_{R}\, \phi)$ and $(e^{i}_{R}\, \phi)$ with the weak and  the
appropriate hyper charge, the anti-Higgs "dresses" correspondingly $(u^{i}_{R}\, anti\phi)$ and 
$(\nu^{i}_{R}\, anti\phi)$. Higgs takes care of the 
masses of the superposition of the weak and hyper charge gauge 
bosons, 
leaving the electromagnetic field massless.
To take care of the masses and mixing matrices of fermions in agreement with the experimental data 
the {\it standard model} postulates  the existence of Yukawa couplings, which are different for different 
family members.} 
\end{table}  

While all the so far observed {\it fermions} are {\it spinors} with the {\it charges in the fundamental 
representations} of the charge groups~\footnote{The "internal degrees" of freedom of particles and 
fields, that means the spin and the charges, are theoretically described by the representations of 
the Lie groups. The same commutation relations of the infinitesimal generators  of the groups 
allow infinite many representations: the scalar one, the fundamental one, the adjoint one, $\cdots$.}
and all the so far observed {\it bosons} are {\it vectors} in the 
{\it adjoint representations with respect to  the charge groups},  the  {\it Higgs} fields 
are {\it scalars} with the {\it charges in the fundamental representations of the 
charge groups}. Therefore, quite a strange object, which reminds us 
of a supersymmetric particle~\footnote{The supersymmetric theories assume that in the low energy regime 
there exist superpartners to the existing particles. The superpartners to existing bosons are fermions 
with the charges in the adjoint representations and the super partners to existing fermions 
are bosons with the charges in the 
fundamental representations.} (but it is not because it does not fit the so called R parity requirement 
for a supersymmetric particle).
The labels "scalar, vector, spinor (fermion)" fields express the behaviour of a field with respect to the 
Lorentz transformations in the space $d=(1+3)$.

The {\em standard model}  never has the ambition to explain its own assumptions, 
leaving the explanation of the open questions to the next step of the theory. 
Although the {\em standard model} leaves many questions unanswered, yet it is, without any doubt, a very 
efficient effective theory: There is so far no experiment which would help to show the next step 
beyond the {\em standard model}, no new  fermions or bosons, no supersymmetric 
particles, even no Higgs yet.

In the literature there are several proposals trying to  go beyond the standard model, most of them 
just extending the ideas of the {\em standard model}, like:
i.)  A tiny extension is the inclusion of the right handed neutrinos into the family (what is done 
in table~\ref{Table I.}). 
ii.) The $SU(3)$ group is assumed to describe -- not explain -- the existence of three families. 
iii.) Like Higgs has the charges in the fundamental representations of the groups, also
Yukawas are assumed to be scalar fields, in the fundamental (fundamental for left handed family members and 
anti-fundamental for the right handed ones) representation of the $SU(3)$ group, belonging to different 
gauge groups for different family members~\cite{Georgi,giudice,belen,ch}.
vi.) Supersymmetric theories assuming the existence of partners to the existing  fermions 
and bosons, with charges in the opposite representations, adjoint for fermions and fundamental for bosons.

The question is: What do the Higgs together with the Yukawa couplings  of the {\it standard model} 
effectively represent? Is the Higgs really a scalar with the fermionic quantum numbers in the charge 
sector, or it is just (so far very efficient) effective  representation for several scalar fields 
which manifest as the Higgs and the  Yukawa couplings?  Are extensions of the Higgs to the Yukawa 
scalar fields with the family charge(s) again in the fundamental representations  
the right way beyond the {\it standard model}?  

To answer any question about the Higgs and the Yukawas one first needs the answer the question:
Where do Yukawas originate, that is where do families originate?

Although effective interactions can have in physics many times quite unexpected shape and yet 
can be very useful (as it is the case, for example, with the by experiments suggested spin-spin 
interaction in several models in the solid state 
physics where the interaction of the electromagnetic origin among many electrons and nuclei involved 
can effectively be expressed with the spin-spin interaction) yet it is hard to accept 
that effective theories of the type where the $SU(3)$ groups  describe  the family 
quantum numbers, with the  scalar dynamical fields which carry the family charges of the fermion kind, 
can make useful predictions for new experiments, where searches depend strongly on the proposed 
theories behind. To my understanding at this stage of physics a new 
more general understanding of fermion and boson fields is needed.

Any new step in theoretical explanation  of the {\it standard model} 
assumptions  must answer the following most urgent open questions: 
\begin{itemize}
\item  What is the origin of families? 
  How many families there are at all?
\item What is the origin of the scalar  fields (the Higgs)? 
 Where do their  masses   (the Higgs mass) and correspondingly  
the masses of the gauge fields originate? 
What is the  origin of the fermion masses,  
 where do Yukawa couplings originate?
\item Where does the dark matter originate?
\end{itemize}
 There are also several other questions which may not be so urgently   answered, like:
Where do dark energy originate?  
What is the origin of charges, and correspondingly of the gauge fields? What does
cause the matter-antimatter asymmetry? And (many) others.

Let me in this discussion demonstrate that the theory unifying spin, charges and families,  called
the {\em spin-charge-family} theory~\cite{norma,pikanorma,NF,gmdn,GN}, looks so far 
very  promising in answering the above and several other open questions, and accordingly offers  
the next step beyond the {\it standard model}.

The {\em spin-charge-family} theory is defined in more than four dimensional space (and is accordingly of 
a Kaluza-Klein type) and therefore not yet acceptable for those who require for all  
theoretical assumptions the existing experimental confirmation. Yet this theory may 
teach us a lot about the open questions of the {\it standard model}, and correspondingly about 
possible origin of families of fermions and of the origin of scalar fields -- the Higgs and Yukawas, 
if taken  as it manifests at low energies.

I shall comment  on the nature of the scalar dynamical fields, with the family (flavour) 
charges  in the fundamental (fermionic) representations~\cite{Georgi,giudice,belen}, extending Higgs 
to Yukawas from the point of view of the {\em spin-charge-family} theory.

\section{Short presentation of the spin-charge-family theory}
\label{mine}

I present here the {\it spin-charge-family} theory from the point of view which shows up 
what new steps beyond the {\it standard model} is the theory offering and in which way.
I present its starting  assumptions and the effective action which the theory manifests 
after several breaks of the starting symmetry (that is after several phase transitions),
manifesting  before the electroweak break massless  families of fermions and massless gauge 
vector bosons with the properties  which the {\it standard model} postulates and are 
presented in tables~\ref{Table I.},~\ref{Table II.}. 
The {\it spin-charge-family} theory  predicts four rather than three so 
far observed families of quarks and leptons at the low energy regime and several scalar 
dynamical fields (the gauge scalar bosons) with  all the charges (with the family quantum 
number included) in the adjoint representations. Effectively, however, these scalar 
fields do behave approximately as the {\it standard model} Higgs and Yukawas. Although 
the detailed calculations are not yet finished, the so far made estimations show that the 
{\it spin-charge-family} theory is in a good way to answer the urgent 
(presented above)  and other open questions of the {\it standard model}.  

The reader is kindly asked to look for more details in the ref.~\cite{NF,AN} and the 
references therein.

Let me start with the main assumption of the {\it spin-charge-family} theory. \\
{\bf i.)} The space has more than $(1+3)$ dimensions. I made a choice of 
$d= (1+13)$\\
{\bf ii.)} One of the two existing Clifford algebra objects, $\gamma^a$, is used to describe 
spin in $d>(1+3)$  of fermions, the other one, I called it $\tilde{\gamma}^a$, to describe families.\\
{\bf iii.)} The simplest action for massless fermions, carrying in $d>(1+3)$  only two kinds of 
the spin, no charges, and for the corresponding gauge fields - 
the vielbeins and the two kinds of the spin connection fields -  in $d> (1+3)$ is taken.\\ 
{\bf iv.)} The breaks of symmetries (phase transitions) are assumed which lead in $d=(1+3)$ 
to the observed phenomena: To the observed massive 
families of quarks and leptons with the observed charges assumed by the {\it standard model}.

Since there exist  two, only two, kinds of the Clifford algebra objects, which 
generate equivalent representations with respect to each other (that is independent spaces), 
and since there exist families of fermions, then if one of these 
two objects is used to describe the spin in $d>(1+3)$ (which then manifest in $d=(1+3)$ as the spin and all 
the charges of one family of fermions), the other has a great chance to properly  describe families.   
Dirac  used 80 years ago $\gamma^a$'s  to describe the spin of fermions enabling the 
success of quantum mechanics. Kaluza-Klein-like theories~\cite{KK}, assuming that space has  more than 
($1+3$) dimensions,  suggest that the spin  together with the angular momentum in higher dimensions
manifests as charges in ($1+3$)~\footnote{In the ref.~\cite{hn} possible ways of solving the 
Witten's~\cite{witten} 
"no-go" theorem for the Kaluza-Klein-type of theories are presented.}. 
As shown in the refs.~\cite{norma,pikanorma,NF}
the second  kind of the Clifford algebra objects has a great chance to properly describe families.

The {\it spin-charge-family} theory starts in $ d= (1 + 13)$  with the simplest possible 
action~\cite{pikanorma,NF} 
which takes into account both kinds of the Clifford operators, $\gamma^a$ and $\tilde{\gamma}^a$ 
(this two kinds of generators anticommute ($\gamma^a$ $\tilde{\gamma}^b$ $+$ $\tilde{\gamma}^b$ $\gamma^a$)
=0),
\begin{eqnarray}
S            \,  &=& \int \; d^dx \; E\;{\mathcal L}_{f} +  
                \int \; d^dx \; E\; (\alpha \,R + \tilde{\alpha} \, \tilde{R}). 
               \end{eqnarray}
The first part describes the fermion degrees of freedom and is just the action for massless fermions
with the two kinds of the spin and no charges interacting correspondingly with  (only) the gravitation 
field -- the vielbeins $f^{\alpha}{}_{a}$ and the two kinds of the spin connection fields, 
the gauge fields of $S^{ab}$ $(= \frac{i}{4} (\gamma^a \gamma^b - \gamma^b \gamma^a))$ and 
$\tilde{S}^{ab}$ $(= \frac{i}{4} (\tilde{\gamma}^a \tilde{\gamma}^b-\tilde{\gamma}^b \tilde{\gamma}^a ))$,
($S^{ab}$ $\tilde{S}^{cd}$ $-$  $\tilde{S}^{cd}$ $S^{ab}=0$ ),
\begin{eqnarray}
\label{wholeaction}
{\mathcal L}_f &=& \frac{1}{2}\, (E\bar{\psi} \, \gamma^a p_{0a} \psi) + h.c.,\quad 
p_{0a }        = f^{\alpha}{}_a p_{0\alpha} + \frac{1}{2E}\, \{ p_{\alpha}, E f^{\alpha}{}_a\}_-, 
\nonumber\\  
   p_{0\alpha} &=&  p_{\alpha}  - 
                    \frac{1}{2}  S^{ab} \omega_{ab \alpha} - 
                    \frac{1}{2}  \tilde{S}^{ab}   \tilde{\omega}_{ab \alpha}.
                    \end{eqnarray}
The Lagrange density for the gauge fields is assumed to be in the starting action linear in the curvature 
$R              =  \frac{1}{2} \, \{ f^{\alpha [ a} f^{\beta b ]} \;(\omega_{a b \alpha, \beta} 
- \omega_{c a \alpha}\,\omega^{c}{}_{b \beta}) \} + h.c. \;,$  
$\tilde{R}      = \frac{1}{2}\,   f^{\alpha [ a} f^{\beta b ]} \;(\tilde{\omega}_{a b \alpha,\beta} - 
\tilde{\omega}_{c a \alpha} \tilde{\omega}^{c}{}_{b \beta}) + h.c.\;.$ 
The action for fermions  manifests after several breaks of symmetries (phase transitions)  in $d=(1+3)$ 
before the electroweak break four  families of   left handed weak charged 
quarks and leptons and  right handed weakless 
(without the weak charge) quarks and leptons~\cite{NF}, just as it is assumed by the {\it standard model}, 
except that an additional $U(1)$ charge exists and that there are also right handed neutrinos, with the 
nonzero value of this  additional $U(1)$ charge and that there are four rather than three families. 
The superposition of the gauge fields $\omega_{ab \alpha}$ and $\tilde{\omega}_{ab \alpha}$ 
manifest after several phase transitions and before the electroweak break as  the known gauge vector 
fields (if $\alpha \in (o,1,2,3)$) and the scalar gauge fields (if $\alpha \in (5,6,7, \cdots),d
$).

While $\bar{\psi} \,E\, \gamma^m p_{0m} \psi, $  with $p_{0m}= p_m - g^a\,\tau^{Ai} A^{Ai}_{m}$,
represents the usual Lagrange density for massless fermions of the {\it standard model}, and 
$p_{0m}$ the covariant momentum with the charges $A$ presented in table~\ref{Table I.}, represents
$\bar{\psi} \,E\, \gamma^s p_{0s} \psi,\, s=7,8; $  the mass term~\cite{NF}: It is the operator $\gamma^s$ 
which transforms all the charges of the right handed members into the charges of the left handed ones, 
doing the job for which the scalar Higgs  with the fermionic charges was postulated. The scalar fields 
appearing in $p_{0s}$
\begin{eqnarray}
\label{mass}
p_{0s}&=& p_s- \sum_{A,i}\, (g^{\tilde A}\, \tau^{\tilde{A}i } \, \tilde{A}^{\tilde{A}i}) -   
(g^Q Q A^{Q}_{s} + g^{Q'} Q' A^{Q'}_{s} + g^{Y'} Y' A^{Y'}_{s})
\end{eqnarray}  
are dynamical scalar fields,  
which in the electroweak phase transition gain a nonzero "vacuum expectation values"\footnote{The "nonzero 
vacuum expectation values" of some fields, which break symmetries, are in the solid state physics, 
well known. Such an example are states of ferromagnetic or anti-ferromagnetic systems.} 
and determine mass matrices of fermions. The two triplet scalar fields $ \tilde{A}^{\tilde{A}i}_{s}$ are  
the gauge fields of $\tilde{\tau}^{\tilde{A}}$ which determine the two $SU(2)$ charges of four 
families, that is the family 
quantum numbers with the properties that each member of four families carries both quantum numbers 
and the massless members transform accordingly as presented in the diagram~\cite{AN,NF}
\begin{equation}
\label{diagramNtaua}
\stackrel{ \stackrel{\tilde{N}^{i}_{L}}{\leftrightarrow} }{\begin{pmatrix} I_{4} & I_{3} \\
 I_{1} & I_{2} \end{pmatrix}} \updownarrow  \tilde{\tau}^{1i}   \quad. 
\end{equation}
$\tilde{N}^{i}_{L}$  are generators of one of the two $SU(2)$ groups  and $\tilde{\tau}^{1i}$ 
the generators of another one ($\tilde{N}^{i}_{L}$ $=\sum_{a,b= 0,1,2,3} \, \tilde{c}^{\tilde{N}^{i}_{L}}_{ab}$ 
$ \tilde{S}^{ab}$,  $\tilde{\tau}^{1i}$ $=\sum_{a,b= 5,6,7,8} \, \tilde{c}c^{\tilde{\tau}^{1i}}_{ab}$ 
$ \tilde{S}^{ab}$). Each family member carries in addition the spin and the charges originating 
in $S^{ab}$: the hyper ($Y=\sum_{ab} c^{Y}_{ab} S^{ab}$), weak $(\tau^{1i}=\sum_{ab} c^{1i}_{ab} S^{ab})$ 
and colour charges $(\tau^{3i}= 
\sum_{ab} c^{3i}_{ab} S^{ab})$~\footnote{In the {\it spin-charge-family} theory the family 
members carry one more $U(1)$ charge,  $Y'= \sum_{ab} c^{Y'}_{ab} S^{ab}$ hyper charge.}. 
In the electroweak break all the scalar fields gaining "nonzero expectation values" determine 
masses and mixing matrices  of fermions.

The calculations done so far show that the operator $\gamma^s$~\footnote{$\gamma^0 \gamma^s\,, s=7,8\,,$ can be formally 
represented as the operator $\sum_{i=1,2} \, |\psi_{L i}><\psi_{R i}|$, $i=1$ stays for the $u$ 
quarks and neutrinos and $i=2$ for the $d$ quarks and electrons, which "rotates" the 
right handed family members into the left handed partners.} and the scalar dynamical 
fields in $p_{0s}$ (all with the charges in the adjoint representations, also with respect to the 
family quantum numbers) manifest~\cite{NF,AN} the  so far observed properties of fermions and gauge fields, 
and explain therefore the role of the Higgs and the Yukawa couplings of the {\it standard model}.

The scalar dynamical fields determine also masses of massive weak boson fields, 
$Z_{m}$  and $W^{\pm}_{m}$. Detailed calculations are in progress and  
the reader is kindly asked to see the refs.~\cite{NF,AN}  and the 
references therein for more information.

\section{Conclusions and predictions of  spin-charge-family theory}
\label{predictions}

The  {\it spin-charge-family} theory predictions are  so far:
\begin{itemize}

\item There are two groups of four families in the low energy regime. The fourth of the lowest 
four families waits to be measured. The fifth family of the higher group of four families is 
stable and it is therefore the candidate to constitute the dark matter.
More about these topics can be found in the refs.~\cite{NF,AN,GN}.

\item The scalar fields which are responsible for mass matrices of fermions (and correspondingly 
for their masses and mixing matrices) have all the charges (with the family charges included) 
in the adjoint representations, that is they behave like so far observed bosons with 
respect to  the established charges and also with respect to the family charges. 
According to the so far made calculations~\cite{NF,AN}  
the {\it spin-charge-family} theory  has a good chance to reproduce the experimental data. 
How these scalar fields can be represented effectively to manifest as the scalar Higgs field 
with the fermion charges and the Yukawa couplings, is explained in the ref.~\cite{NF}.

\item It is evident from the {\it spin-charge-family} theory that besides the known gauge 
fields also the scalar fields are the interaction fields. Accordingly, since they are effectively 
representing  the  Higgs and the Yukawa couplings, also the {\it standard model} Yukawas 
are the interacting fields. 

\item Searching for scalar fields will manifest that there are several 
dynamical scalar fields,  not just one, and that each of them couples differently to different 
family members.

More about this topic can be read in the ref.~\cite{NF}. The calculations are under 
consideration~\cite{AN}.

\item There are no supersymmetric partners of the so far observed fermion and boson fields, at least 
not in the low energy regime.

\item The extensions of the {\it standard model}~\cite{Georgi,giudice,belen}, where also Yukawas are 
taken as dynamical fields with the charges in the fundamental or anti-fundamental representations 
of several $SU(3)$ groups, distinguishing among the left handed quarks and leptons 
and among all the right handed family members (not quite because of 
the right handed neutrinos missing in almost all these kinds of models)  can, according to the {\it 
spin-charge-family} theory hardly be of some help in predicting future experiments. 
While the {\it standard model} Higgs is, together with the Yukawa couplings, a simple and so 
far very efficient replacement for the dynamical scalar field predicted by the 
{\it spin-charge-family} theory, the extensions~\cite{Georgi,giudice,belen} start to be  
complicated. The experiences in the nuclear physics and solid state physics speak against such models.

\end{itemize}

There are also several other predictions, not yet enough studied to be commented here.

Let me conclude this paper with the statement: The {\it spin-charge-family} theory does have the answers 
to open questions, which are to my understanding the most urgent to be answered for any new successful step 
beyond the {\it standard model}. I doubt that trying to explain only one of the "urgent 
open questions" (presented above) can bring much new insight into the assumptions of the 
{\it standard model}. 

The {\it spin-charge-family} theory offers not only the answer to the question why  we have 
more than one family, and how many there are, it explains also the origin of the Higgs and the 
Yukawa couplings, of the charges and the gauge fields. 


According to these predictions there is no supersymmetric particles at the low energy regime.
But, without doubts, there are several additional dynamical fields - interactions.

Work is in progress and should show (the calculations done so far are  promissing) 
that although on the tree level the mass matrices of quarks and leptons are quite far from leading  
to the observed properties of quarks and leptons,  loop corrections change them so that the 
results for the lower three families agree with the experimental data.

The question is what if the {\it spin-charge-family} theory is not  what the nature has chosen?
Even in this case the theory is teaching us how to make the models with the scalar dynamical 
fields which behave as bosons and which have a chance to answer all the urgent open questions.

\end{document}